\documentclass[%
 reprint,
 amsmath,amssymb,
 aps,
]{revtex4-1}

\usepackage{graphicx}
\usepackage{dcolumn}
\usepackage{bm}

\begin{document}


\title{Charge-density-wave phase, mottness and ferromagnetism in monolayer $1T$-NbSe$_2$} 
\author{Diego Pasquier}
\affiliation{Institute of Physics, Ecole Polytechnique F\'{e}d\'{e}rale de Lausanne (EPFL), CH-1015 Lausanne, Switzerland}
\author{Oleg V. Yazyev}%
 \email{E-mail: oleg.yazyev@epfl.ch}
\affiliation{Institute of Physics, Ecole Polytechnique F\'{e}d\'{e}rale de Lausanne (EPFL), CH-1015 Lausanne, Switzerland}

\begin{abstract}
The recently investigated $1T$-polymorph of monolayer NbSe$_2$ revealed an insulating behaviour suggesting a star-of-David phase with $\sqrt{13}\,\times\sqrt{13}$ periodicity associated with a Mott insulator, reminiscent of $1T$-TaS$_2$.
In this work, we examine this novel two-dimensional material from first principles.
We find an instability towards the formation of an incommensurate charge-density-wave (CDW) and establish the star-of-David phase as the most stable commensurate CDW.
The mottness in the star-of-David phase is confirmed and studied at various levels of theory: the spin-polarized generalized gradient approximation (GGA) and its extension involving the on-site Coulomb repulsion (GGA+$U$), as well as the dynamical mean-field theory (DMFT).
Finally, we estimate Heisenberg exchange couplings in this material and find a weak nearest-neighbour ferromagnetic coupling, at odds with most Mott insulators. 
We point out the close resemblance between this star-of-David phase and flat-band ferromagnetism models.
\end{abstract}

\date{\today}

%
\maketitle

\section{Introduction}
Transition metal dichalcogenides (TMDs) have been extensively studied for their charge-density-wave phases \cite{wilson_charge-density_1974, wilson_charge-density_1975, castro_neto_charge_2001, rossnagel2011origin}, historically being the first materials where the Peierls instability \cite{peierls1955quantum} manisfests itself, although this point of view has been frequently challenged in the last few years \cite{johannes_fermi_2008, zhu_classification_2015}.
More recently, TMDs (for recent reviews see e.g. \cite{chhowalla_chemistry_2013, manzeli_2d_2017}) have further attracted attention due to their novel topological properties \cite{qian_quantum_2014, soluyanov_type-ii_2015}, unconventional Ising superconductivity \cite{lu_evidence_2015, xi_ising_2016, yuan_possible_2014, saito_superconductivity_2016}, as well as the possibility to thin them down to a single layer \cite{novoselov_two-dimensional_2005}, leading to a rich family of two-dimensional (2D) materials that includes semiconductors with promising technological applications \cite{wang2012electronics}.

TMDs with chemical composition $MX_2$ are layered materials, each layer consisting of a transition metal ($M$ = Ti, V, Nb, Ta, etc.) forming a triangular lattice sandwiched between two atomic planes of chalcogen atoms ($X$ = S, Se, Te).
The local coordination sphere of the transition metal can have either trigonal prismatic or distorted octahedral symmetry, giving rise to two families of polytopes, referred to as $2H$ and $1T$, respectively, where $1$ and $2$ stand for the number of inequivalent layers in the unit cell for bulk materials.
The different coordination environments lead to distinct crystal field splittings of the $d$-like bands and therefore very different electronic properties \cite{chhowalla_chemistry_2013}.

Among all the TMDs, $1T$-TaS$_2$ displays arguably the most complex phase diagram.
Indeed, $1T$-TaS$_2$  exhibits a series of structural phase transitions, that involves one second-order and two first-order transitions, upon decreasing temperature \cite{wilson_charge-density_1975, di_salvo_low_1977, sipos_mott_2008}.
The low-temperature commensurate CDW phase is characterized by the formation of star-of-David clusters of Ta atoms in a $\sqrt{13}\times\sqrt{13}$ supercell associated with the emergence of a narrow band crossing the Fermi level \cite{darancet_three-dimensional_2014, ritschel_orbital_2015}, favouring the opening of a Mott correlation gap.
Moreover, it has recently been pointed out that no trace of magnetic order is observed down to very low temperatures, indicating a possible quantum spin liquid (QSL) state \cite{law_1t-tas2_2017}.

Only known so far in the $2H$ phase, $1T$-NbSe$_2$ has recently been successfully synthesized in a monolayer form \cite{nakata_monolayer_2016}.
Niobium is situated in the same column of the periodic table as Tantalum, which implies that these two transition metal elements are isoelectronic and have formal $d$-shell populations of $4d^1$ and $5d^1$ in NbSe$_2$ and TaS$_2$, respectively.
It has been found that a superlattice is formed in monolayer $1T$-NbSe$_2$ and that the electronic structure exhibits an insulating energy gap of $\sim0.4$ eV, strongly suggesting a phase diagram analogous to $1T$-TaS$_2$.

The purpose of this paper is to provide a first-principles study of this new material, including the instability of the metallic undistorted $1T$ phase towards a CDW phase, structural properties and different scenarios for the nature of the gap, correlation effects and magnetism.
Our work confirms the $\sqrt{13}\,\times\sqrt{13}$ phase as the most stable commensurate CDW phase as well as the opening of a correlation gap that is to some extent captured even by spin-polarized GGA calculations. GGA+$U$ and GGA+DMFT calculations provide further insight and suggest a gap of the charge transfer type.
An estimation of Heisenberg exchange couplings surprisingly indicates a ferromagnetic ground state, contrary to what one would expect in a Mott insulator. 
We suggest that, if confirmed, the ferromagnetism strongly resembles the flat-band ferromagnetism  \cite{mielke_ferromagnetic_1991, mielke_exact_1992, mielke_ferromagnetism_1993, tasaki_ferromagnetism_1992, tasaki_nagaokas_1998, ichimura_flat-band_1998, watanabe_theoretical_1997, penc_ferromagnetism_1996} effect in multiband Hubbard models and that this star-of-David phase could be a real material realization of this effect in 2D.

This paper is organized as follows.
Section \ref{sec:method} briefly describes the computational methodology.
In Section \ref{sec:cdw}, we study the fermiology and the phonon dispersion of the undistorted $1T$ phase, as well as possible commensurate superlattices.
In Sections \ref{sec:mott} and \ref{sec:magnetism}, we present an analysis of the electronic structure and magnetism of the $\sqrt{13}\,\times\sqrt{13}$ phase.
Section \ref{sec:conclusion} offers conclusions.

\section{Computational methodology}\label{sec:method}
First-principles density functional theory (DFT) calculations were performed using the \textsc{Quantum ESPRESSO} package \cite{giannozzi2009quantum}. 
The interaction between the valence and core electrons is described by means of ultrasoft pseudopotentials \cite{vanderbilt_soft_1990} (available from the \textsc{pslibrary} \cite{dal_corso_pseudopotentials_2014, pslibrary}), explicitly including the $s$ and $p$ semi-core electrons as valence electrons for Nb atoms.
The plane-wave cutoffs are set to $60$ and $300$~Ry for the wave functions and charge density, respectively.
The exchange-correlation functional is approximated by the generalized gradient approximation according to Perdew, Burke and Ernzerhof (PBE) \cite{perdew_generalized_1996}. For GGA+$U$ calculations, we adopt the simplified formulation of Cococcioni and de Gironcoli \cite{cococcioni_linear_2005}, with a Hubbard parameter $U=3.02$~eV for Nb $4d$ orbitals, calculated from linear response in a supercell of the undistorted $1T$ phase containing $75$ atoms.
Brillouin zone integration is performed on a $24\times24\times1$ k-points mesh ($8\times8\times1$ and $6\times6\times1$ for the $\sqrt{13}\,\times\sqrt{13}$ and $4\times4$ supercells, respectively) and a Marzari-Vanderbilt smearing \cite{marzari_thermal_1999} of $1$~mRy.
To simulate the monolayer form, we include approximatively $13~\, \text{\AA}$ of vacuum between periodic replicas.
Lattice constants and atomic positions of various phases are determined by fully relaxing the structure at the PBE level until all the Hellmann-Feynman forces are less than $10^{-4}$~Ry/Bohr.  
The spin-orbit coupling is not included but its effect is described in the Supplementary information document \cite{suppl}.

The phonon dispersion is calculated within density functional perturbation theory (DFPT) \cite{baroni_phonons_2001}, using a denser mesh of $84\times84\times1$ k-points and a larger smearing of $5$~mRy.
To plot the full dispersion, we have calculated the phonons on a $12\times 12$ grid of q-points and used Fourier interpolation.
In addition, we have computed the dispersion close to the CDW wave vector by performing a DFPT calculation for several points in its vicinity, using different smearings of $10$ mRy, $5$ mRy and $2.5$ mRy and a denser grid of $192\times 192$ k-points to ensure convergence of the imaginary frequencies. 

Dynamical mean-field theory (DMFT) \cite{georges_dynamical_1996} calculations are performed using the \textsc{AMULET} code \cite{amulet}.
The quantum impurity problem is solved with the continuous-time quantum Monte-Carlo (CT-QMC) algorithm \cite{gull_continuous-time_2011}  with ten millions QMC steps.
The simplified fully localized limit prescription is adopted to account for double counting.
The spectrum is obtained with the maximum entropy method.

Maximally localized Wannier functions (MLWF) \cite{marzari_maximally_1997, marzari_maximally_2012} are obtained using the \textsc{Wannier90} code \cite{mostofi_updated_2014}.
The susceptibility is calculated on a dense $400\times400\times1$ k-points grid with Wannier-interpolated bands.

\section{Charge-density-wave phases}\label{sec:cdw}
We begin our discussion by determining the structural and electronic properties of the undistorted $1T$ polymorph of monolayer NbSe$_2$.
The latter contains three atoms per unit cell and belongs to the symmorphic $D^3_{3d}$ space group.
The lattice constant and the Nb$-$Se distance at the PBE level are $a=3.49 \, \text{\AA}$ and $d_{\text{Nb-Se}}=2.62 \, \text{\AA}$, respectively.

\begin{figure}
\includegraphics[width=8.6cm]{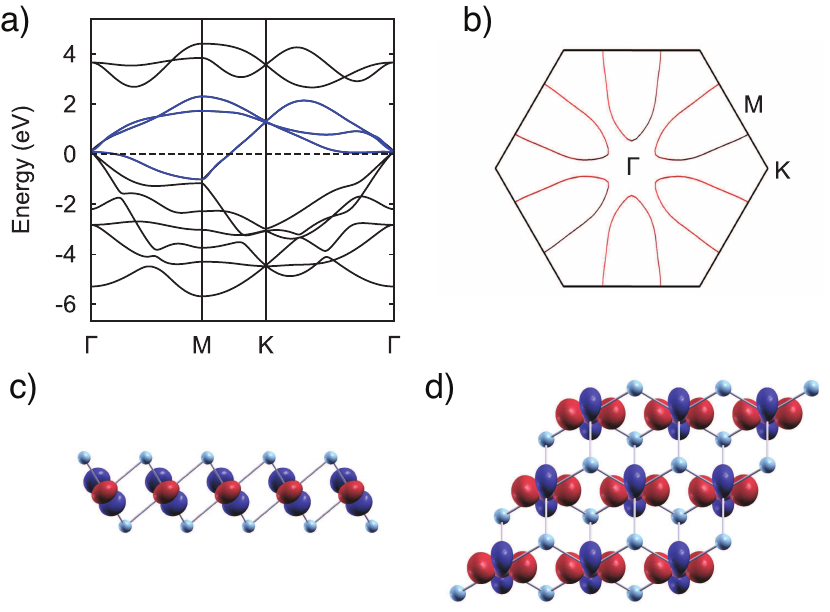}
\caption{\label{fig1} (a) GGA band structure of the undistorted $1T$-phase of monolayer NbSe$_2$. The t$_{2g}$ bands are emphasized in blue. The dashed line corresponds to the Fermi energy, set to zero. (b) Fermi surface of monolayer NbSe$_2$. (c),(d) Ball-and-stick representation of the undistorted $1T$ phase of monolayer NbSe$_2$ with an isosurface plot of one of the three symmetry-equivalent t$_{2g}$-like Wannier functions. Selenium atoms are shown in blue.}
\end{figure}

The electronic structure and the t$_{2g}$ Fermi surface are shown in Figure~\ref{fig1}.
Since the spin-orbit coupling does not play an important role, we neglect it but briefly describe its effect in the Supplementary material \cite{suppl}.
The three t$_{2g}$ bands are filled with one electron.
The bandwidth is rather large ($\sim 3$ eV), implying that the moderate electron-electron interactions can be neglected at this point.
On the other hand, the t$_{2g}$ electrons are prone to form $\sigma$-bonds due to their directional character, implying a large coupling to a local bond-stretching phonon. 
The latter is, to the best of our understanding, responsible for the recurrent occurrence of CDWs in the $1T$ dichalcogenides and lead to stronger distortions when the filling is closer to half-filling, as e.g. in $1T'$-MoS$_2$ \cite{pizzochero2017point} or ReS$_2$ \cite{tongay2014monolayer}, in which strong metal-metal bonds are formed.
The Fermi surface is typical of group V $1T$ dichalcogenides and displays pseudo-nesting, favouring density wave instabilities with incommensurate wave vectors $\mathbf{Q}_i=Q_{\mathrm{ICDW}}\mathbf{b}_i$, where $\mathbf{b}_i$ ($i=1, \,2, \,3$) are the three reciprocal lattice vectors of a triangular lattice and $Q_{\mathrm{ICDW}} \approx 0.25-0.33$ \cite{wilson_charge-density_1975, woolley_band_1977}, depending on material-dependent details of the electronic structure.
%

\begin{figure*}[t]
\includegraphics[width=17cm]{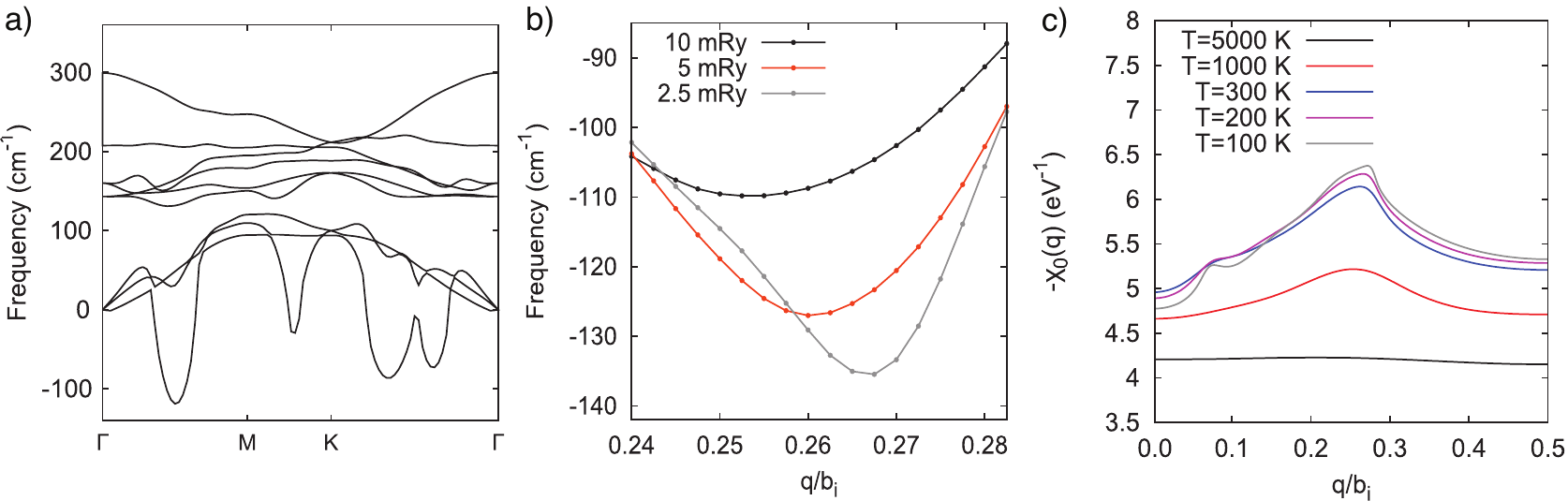}
\caption{\label{fig2} (a) Calculated phonon dispersion for the undistorted $1T$ phase of monolayer NbSe$_2$ obtained by Fourier interpolation. Imaginary frequencies are plotted as negative. (b) Phonon dispersion close to the wave vector of maximum softening with different electronic population smearing values. Each point corresponds to a DFPT calculation. (c) Calculated bare static susceptibility along the $\Gamma-M$ direction at different electronic temperatures.}
\end{figure*}

Figure~\ref{fig2} shows the calculated phonon dispersion curves and bare static susceptibility along the $\Gamma-M$ direction.
Neglecting matrix elements, the susceptibility reads 
\begin{equation}
\chi_0(q)= \frac{1}{N_k}\sum_{k,n,m}\frac{f(\epsilon_{n,k+q})-f(\epsilon_{m,k})}{\epsilon_{n,k+q}-\epsilon_{m,k}} \, \, ,
\end{equation}
where $f(\epsilon_{n,k})$ is the Fermi-Dirac distribution and $\epsilon_{n,k}$ are the Kohn-Sham energies.
The susceptibility is proportional to the phonon self-energy in the random phase approximation, favouring soft phonon modes when it is enhanced at a particular wave vector \cite{chan1973spin}.
One can see that the system is unstable against the formation of a CDW with momentum $Q_{\mathrm{ICDW}}\approx 0.26$, corresponding to the maximum of the susceptibility at $T=300$~K (Fig.~\ref{fig2}b).
At lower temperatures, the maximum is shifted closer to $Q_{\mathrm{ICDW}} = 0.27$.
Accordingly, the calculated phonon softening becomes stronger closer to  $Q_{\mathrm{ICDW}} = 0.27$ when a smaller smearing is used.
We also observe that at $T = 5000$~K the susceptibility is completely flat as the Fermi surface is blurred. 
The incommensurability of the soft phonon mode and its correlation with the maximum of the susceptibility demonstrate the effect of the fermiology on the CDW (Fig.~\ref{fig2}c), even if we stress that the latter is possible only in the presence of a rather strong electron-phonon coupling due to imperfect nesting.

As understood by McMillan \cite{mcmillan_landau_1975, mcmillan_theory_1976}, density waves can further gain energy by adopting a commensurate periodicity characterized by a momentum $\mathbf{Q_{\mathrm{CCDW}}}$ close to $\mathbf{Q_{\mathrm{ICDW}}}$.
This can lead to first-order incommensurate-to-commensurate phase transitions (lock-in transitions) as the temperature is lowered.
Such transitions come from higher-order terms of the free energy and are therefore not captured by a phonon calculation.
The calculated $Q_{\mathrm{ICDW}}\approx 0.26$ suggests either $4\times4$ or $\sqrt{13}\times\sqrt{13}$ periodicity. 
In the latter case, each unit cell contains an odd number of electrons and an insulating gap, as observed in experiments, can only come from electron correlations.
On the other hand, the $4\times4$ cell could possibly be a normal band insulator.
We have therefore addressed both scenarios by relaxing atomic positions (starting from randomized ones) and lattice vectors in the two supercells.

For the $4\times4$ cell, we obtain an energy gain of $49$~meV per NbSe$_2$ formula unit compared to the undistorted $1T$ phase and a magnetically ordered metallic phase (see supplementary information \cite{suppl}), whereas for the $\sqrt{13}\times\sqrt{13}$ cell we obtain the star-of-David phase with a larger energy gain of $69$ meV/f.u. and a Mott insulator phase (see next section). Another possibility would be that the CDW remains incommensurate down to zero temperature.
However, incommensurate CDWs in the dichalcogenides usually have a rather small effect on the electronic structure so that it is unlikely that a gap of $\sim 0.4$ eV could be opened.

\section{Mottness in the Star-of-David phase}\label{sec:mott}

\begin{figure*}[t]
\includegraphics[width=17cm]{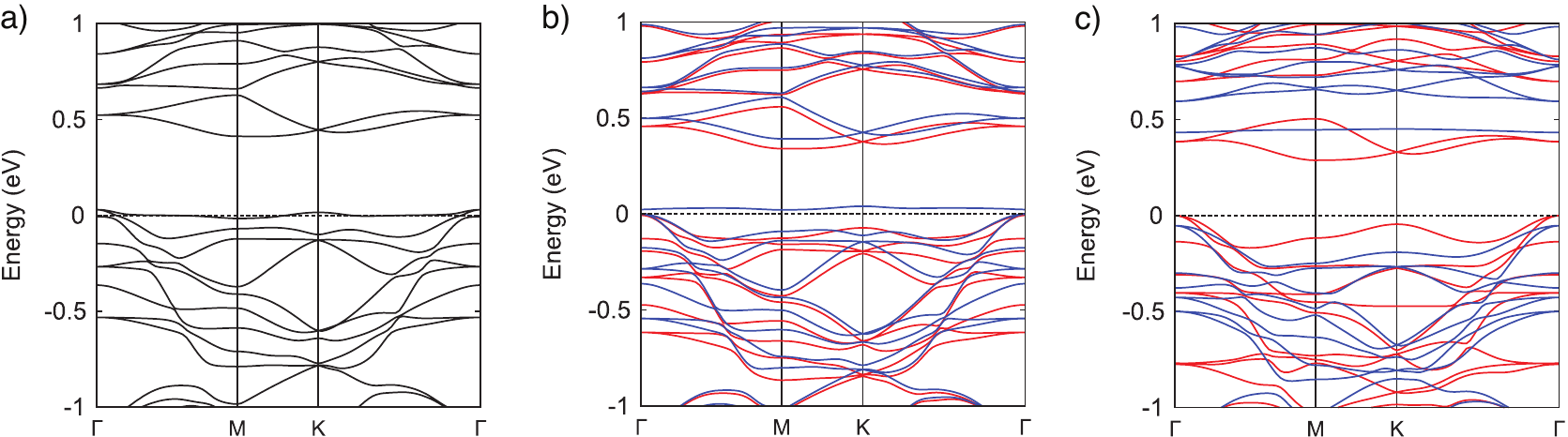}
\caption{\label{fig3} (a) Electronic band structure of the $\sqrt{13}\,\times\sqrt{13}$ CCDW phase of monolayer NbSe$_2$ obtained from non-spin-polarized GGA calculations. The dashed line corresponds to the Fermi level, set to zero. (b),(c) Electronic band structures obtained from spin-polarized (b) GGA and (c)  GGA+$U$ ($U=3.02$~eV) calculations. The up and down spin bands are shown in red and blue, respectively.}
\end{figure*}

We now proceed to study the electronic structure of the star-of-David phase at various levels of theory.
%
As one can see in Figure~\ref{fig3}a, a very narrow band crossing the Fermi level emerges in the GGA band structure. 
Spin-polarized GGA already captures some correlation effects and can sometimes describe mottness approximatively (but not in quantitative agreement with experiments \cite{anisimov_band_1991}), together with a magnetic solution.
In Figure~\ref{fig3}b, we observe a small band gap of $\sim 20$~meV at the spin-polarized GGA level with a total magnetic moment of $1$~$\mu_B$ per supercell that contains one David star. 
The computed gap is clearly too small compared to the experiments, therefore we add an on-site Hubbard repulsion $U = 3.02$~eV for Nb $4d$ orbitals.
The calculated gap is now $\sim 0.3$ eV (Fig.~\ref{fig3}c), in better agreement with the experimental data. 
However, the gap appears to between the "uncorrelated" bands rather than between the lower Hubbard band (LHB) and the upper Hubbard band (UHB), as expected in Ref.~\onlinecite{nakata_monolayer_2016}. We note that the flat LHB and UHB bands can still be distinguished amongst the "uncorrelated" bands in Fig.~\ref{fig3}c.

\begin{figure*}
\includegraphics[width=17cm]{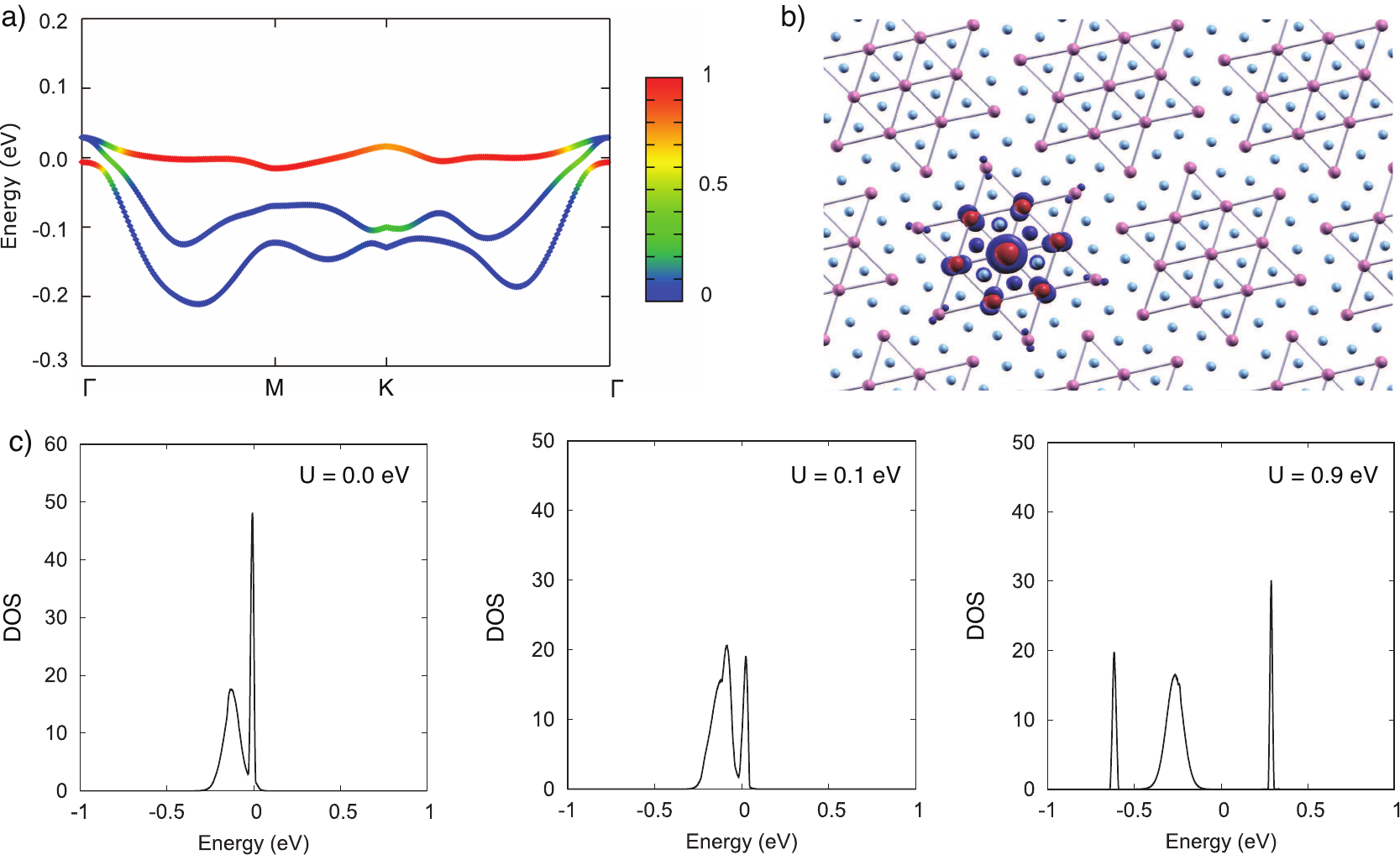}
\caption{\label{fig4} (a) Three-band model chosen for the DMFT calculations with the orbital weight of the type I Wannier function color-coded. (b) Ball-and-stick representation of the star-of-David phase with an isovalue plot of the type I Wannier function. Nb$-$Nb bonds are drawn to facilitate the visualization.  (c) Spectra obtained by analytic continuation of the imaginary-time Green's functions for  $U = 0.0$, $0.1$ and $0.9$~eV. The Fermi energy is set to zero.}
\end{figure*}

To gain further insight, we derive a minimal three-bands (occupied by five electrons) tight-binding model in the basis of maximally localized Wannier functions. 
We obtain, as can be seen on Figure~\ref{fig4}, one Wannier function (type I WF) localized at the center of the star with a spread of $22 \, \text{\AA}^2$, giving rise to the narrow band and two Wannier functions (type II WFs, see Supplementary information \cite{suppl}) with larger spreads and more weights on the outer Nb atoms of the David stars, hybridizing very weakly with the type I WF.
This choice of model allows to capture the bands crossing the Fermi level and to disentangle the narrow "correlated" band, constituting therefore a minimal model to understand the opening of a correlation gap. 
Treating only the type I WF as correlated with a single variable on-site Hubbard parameter $U$, we solve the model with DMFT in the paramagnetic phase with an inverse temperature of $40$~eV$^{-1}$ ($T \approx 300$ K). 

%

Since the band derived from type I WFs is nearly flat with a bandwidth of $\sim 30$~meV, it splits into a LHB and a UHB upon any small interaction, explaining why the GGA functional can already capture the gap opening.
With a sufficiently large Hubbard $U$, a gap opens between the type II bands and the UHB (charge transfer insulator) and the orbital population of the type I WF changes from $1.18$ in GGA to $1.0$ in GGA+DMFT. 
A Hubbard parameter $U \sim 0.9$~eV gives a gap between the type II bands and the UHB consistent with the GGA+$U$ calculation. 
Obviously, the $U$ parameter for the type I WF is expected to be smaller than that for the Nb $4d$ orbitals in GGA+$U$ due to a larger spread. 

We note that in the GGA+$U$ band structure (Fig.~\ref{fig3}c), while a flat UHB is easily recognizable, the LHB appears to further hybridize with other bands, even if a flat-like band is seen at $\sim 0.5$ eV below the valence band maximum. 
This suggests that it would be interesting to compare this minimal three-bands models with more elaborate models containing more bands and to take into account charge self-consistency, but this is beyond the scope of the present work. 

\section{Magnetic phases}\label{sec:magnetism}
In Mott insulators, the low-energy degrees of freedom are localized spins whose interactions lead to long-range magnetic order below a characteristic temperature, unless prevented by strong fluctuations (i.e. a QSL state).
It is therefore natural to study the mean-field magnetic solutions obtained from DFT to anticipate the character of magnetic correlations expected in a material.

\begin{figure}[t]
\includegraphics[width=8cm]{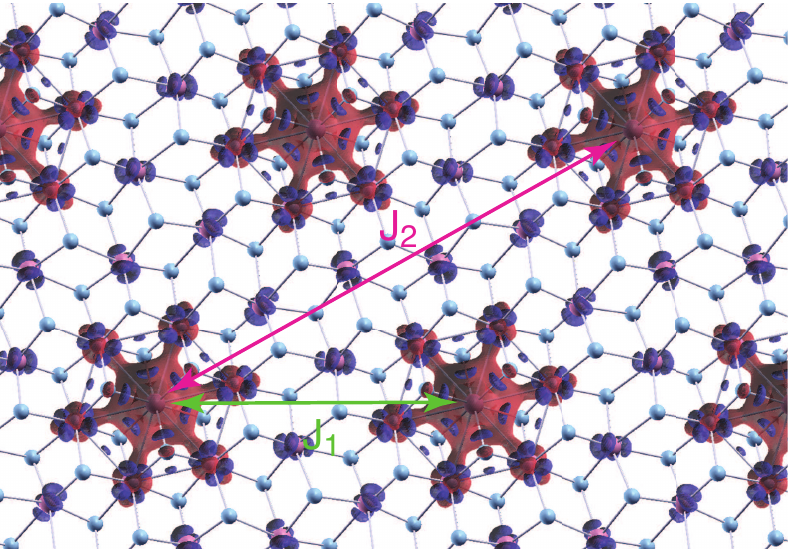}
\caption{\label{fig5} Spin polarization density in monolayer NbSe$_2$ obtained at the GGA+$U$ level in the ferromagnetic phase. A small isovalue of $0.0025$~$a_0^{-3}$ was chosen  to visualize the opposite polarization on the outer star-of-David atoms. The definitions of nearest-neighbour ($J_1$) and next-nearest-neighbour ($J_2$) exchange coupling are indicated.}
\end{figure}

In Figure \ref{fig5}, we present an isovalue plot of the spin polarization density obtained from the GGA+$U$ calculations.
While the total magnetic moment is $1$ $\mu_B$ per star-of-David ($S=1/2$ Mott insulator), the absolute magnetization is found close to $3 \mu_B$/star. 
This is an effect of the on-site Hubbard repulsion, since in the GGA case, the latter is close to one ($1.19 \, \mu_B$/star).
In the GGA+$U$ solution, the Nb atom at the center of the star acquires a larger magnetic moment ($0.8$ against $0.2 \, \mu_B$), while its six nearest-neighbours Se atoms, as well as the six outer Nb atoms, acquire small opposite magnetic moments, as can be seen in the spin polarization plot.
Our GGA+$U$ solution therefore bears resemblance with ferrimagnetism.
However, we stress that the opposite magnetic moments are the consequence of a spin-splitting of the lower bands induced by the magnetic moment associated with the LHB in GGA+$U$. 
Focusing on the global properties of the system, we address the question whether the total spins on neighbouring stars couple ferromagnetically or antiferromagnetically \cite{anisimov_band_1991}.
We therefore compare the total energies of different spin configurations in the $2\sqrt{13}\times\sqrt{13}$ and $\sqrt{3}\sqrt{13}\times\sqrt{3}\sqrt{13}$ supercells (containing two and three stars per supercell, respectively) to extract effective nearest-neighbour and next-nearest-neighbour Heisenberg exchange couplings $J_1$ and $J_2$, as illustrated in Figure \ref{fig5}, assuming that further couplings can be neglected.
We stress that we are aware that DFT can sometimes give misleading results for magnetic properties, but more accurate wave functions method would be prohibitive for this system and we therefore restrict ourselves to GGA and GGA+$U$.

\begin{table}
\caption{\label{tab:heisenberg} Calculated nearest-neighbour ($J_1$) and next-nearest-neighbour ($J_2$) ferromagnetic exchange couplings in Kelvins.}
\begin{ruledtabular}
\begin{tabular}{lll}
 & $J_1$ (K) & $J_2$ (K)\\
\hline\\
GGA & 2.38 & 0.12\\
GGA+$U$ & 4.77 & 0.04\\
\end{tabular}
\end{ruledtabular}
\end{table}

The estimated magnetic exchange couplings are reported in Table~\ref{tab:heisenberg}. 
We find a weak nearest-neighbour ferromagnetic coupling and a negligible next-nearest-neighbour coupling.
This is rather unexpected since Mott insulators are usually antiferromagnetic, with a few exceptions such as YTiO$_3$ \cite{itoh1999orbital} or Ba$_2$NaOs$_6$ \cite{erickson_ferromagnetism_2007}.
We have also verified that introducing the spin-orbit coupling does not affect the sign of the magnetic exchange coupling parameters, even though it gives rise to small anisotropies (see the supplementary information \cite{suppl}).
A possible scenario for the occurrence of ferromagnetism in multiband Hubbard models is the so-called flat-band ferromagnetism studied by Mielke and Tasaki \cite{mielke_ferromagnetic_1991, mielke_exact_1992, mielke_ferromagnetism_1993, tasaki_ferromagnetism_1992, tasaki_nagaokas_1998}. 
Flat-band ferromagnetism can emerge, for instance, on the Kagome lattice with nearest-neighbour hoppings only \cite{pollmann_kinetic_2008, tanaka_stability_2003, chisnell_topological_2015}. 
While a perfectly flat band requires fine-tuning of the model parameters unlikely to happen in any real material, ferromagnetism is robust against some deviations \cite{penc_ferromagnetism_1996, tanaka_stability_2003} if the (nearly) flat-band is at half-filling.
In the monolayer $1T$-NbSe$_2$ case, the flat-band has some dispersion and overlaps in energy with two other bands.
Intuitively, the direct antiferromagnetic exchange is expected to be small because the correlated type I Wannier function are at the center of the stars and have hence small direct hoppings.
Therefore, higher-order processes can become dominant and ferromagnetic couplings can be enabled depending on the sign of the different hopping parameters.
It is expected that several mechanisms are involved, including the effect of the spin polarization of the ``uncorrelated'' bands, and that a quantitative model would likely be rather complicated.

We point out that monolayer $1T$-TaS$_2$ seems even closer to the ideal flat-band model since the narrow band is well isolated. 
We have verified that in this system the magnetic exchange coupling is also ferromagnetic at the GGA and GGA+$U$ levels of theory (in agreement with Ref.~\onlinecite{yu_electronic_2017}). 
We stress that this is not in contradiction with the absence of magnetism observed experimentally, since all experimental studies of magnetism so far were carried out on bulk materials, for which both experiments and calculations suggest significant dispersion between the layers and the existence of a Fermi surface \cite{darancet_three-dimensional_2014, ngankeu_quasi-one-dimensional_2017, ribak_gapless_2017, yu_electronic_2017}.
On the other hand, the ferromagnetic scenario does not seem to agree with the recent proposal of a quasi-2D quantum spin liquid phase in $1T$-TaS$_2$ \cite{law_1t-tas2_2017}, that could occur, e.g. in a $J_1$-$J_2$ antiferromagnetic model on a triangular lattice with $0.08 \leq J_2/J_1 \leq 0.16$ \cite{iqbal_spin_2016}. 
It would therefore be interesting to address experimentally the possible magnetic ordering in monolayer $1T$-NbSe$_2$ and $1T$-TaS$_2$ at low temperatures.

\section{Conclusions}\label{sec:conclusion}
In our work, we addressed by means of first-principles calculations monolayer $1T$-NbSe$_2$ that was recently realized experimentally.
We found an instability against an incommensurate CDW and established the $\sqrt{13}\times\sqrt{13}$ CCDW with the star-of-David distortion as the most stable phase.
Our calculations performed at the level of DFT, DFT+$U$ and DMFT identify this configuration as a Mott insulator.
Finally, we suggested the possible existence of ferromagnetic ordering in this star-of-David phase and pointed out the resemblance with the so-called flat-band ferromagnetism scenario.
The emergence of the narrow band close to the Fermi level in the CCDW phase leads to exotic physics making these materials unique in the family of the TMDs. \\

\section*{Acknowledgements}
We acknowledge funding by the European Commission under the Graphene Flagship (Grant agreement No.~696656). 
We thank Hyungjun Lee, QuanSheng Wu and Vamshi Katukuri for useful discussions. First-principles calculations were performed at the Swiss National Supercomputing Centre (CSCS) under project s675 and the facilities of Scientific IT and Application Support Center of EPFL.

\vskip 0.5cm
\noindent
\textit{Note:} While the manuscript was in the final stage of preparation, a related work on monolayer $1T$-NbSe$_2$ has been reported \cite{Calandra}.
\bibliographystyle{apsrev4-1}
\bibliography{nbse2}
\end{document}